\documentclass[epj,twocolumn,preprintnumbers,nopacs]{svjour}
\pdfoutput=1 
\usepackage{epsfig}
\usepackage{graphicx}
\usepackage{dcolumn}
\usepackage{bm}
\usepackage{amssymb}
\usepackage{subfigure}
\usepackage{mathrsfs}
\usepackage{graphicx}
\usepackage{multirow}
\usepackage{rotating}
\usepackage{url}
\usepackage{color}
\usepackage{setspace}
\usepackage{amsmath}
\usepackage{textpos}
\usepackage{hyperref}

\titlerunning{Search for Massive Colored Scalars in Four-Jet Final States with ATLAS}
\newcommand{\sgluontitle}{\begin{boldmath} Search for Massive Colored Scalars in Four-Jet Final States in $\sqrt{s}=$7~TeV proton-proton collisions with the ATLAS Detector \end{boldmath}}

\def\GeV{\ifmmode {\mathrm{\ Ge\kern -0.1em V}}\else
                   \textrm{Ge\kern -0.1em V}\fi}%
\def\TeV{\ifmmode {\mathrm{\ Te\kern -0.1em V}}\else
                   \textrm{Te\kern -0.1em V}\fi}%
\def\ipb{\mbox{pb$^{-1}$}}
\def\Wboson{\ensuremath{W}}%
\def\antibar#1{\ensuremath{#1\bar{#1}}}
\def\ttbar{\antibar{t}}

\begin{document}
\title{\sgluontitle \begin{textblock}{8}[0,1](0,-2.8){\small CERN-PH-EP-2011-154, Submitted to EPJC Letters}
\end{textblock}
}

\author{The ATLAS Collaboration}
\institute{}
\date{}

\abstract{
A search for pair-produced scalar particles decaying to a four-jet final state is presented. 
The analysis is performed using an integrated luminosity of 34 \ipb\ recorded by the ATLAS detector in 2010. 
No deviation from the Standard Model is observed. For a scalar mass of 100 GeV (190~GeV) the limit on the 
scalar gluon pair production cross section at 95\% confidence level is 1~nb (0.28~nb). 
When these results are interpreted as mass limits, 
scalar-gluons (hyperpions) with masses of 100 to 185~\GeV\ (100 to 155~\GeV) are excluded at 95\% confidence level
with the exception of a mass window of width about 5~\GeV\ (15~\GeV) around 140~\GeV. 
}

\maketitle

At hadron colliders, the search for new phenomena in fully hadronic final states without missing transverse energy or leptons is experimentally challenging because of the large multijet background. Recent studies used the dijet mass spectrum and the dijet angular distribution observed at the 
LHC~\cite{Chatrchyan2011123,Aad:2011aj,Collaboration:2011fq} to search for physics beyond the Standard Model.
Some extensions of the Standard Model predict new phenomena in events with higher jet multiplicity.
In the six-jet final state CDF~\cite{Aaltonen:2011sg} and CMS~\cite{Chatrchyan:2011cj} have excluded 
R--parity violating gluinos, the supersymmetric partners of the gluons, with masses from 200~GeV to 280~\GeV~using a model-independent search.
This letter describes a search for pair-produced scalar particles decaying to two jets, 
leading to a four-jet final state with two jet-jet resonances and no missing transverse energy.

Two scenarios serve as a guideline and motivation for the analysis: 
the extension of the Standard Model with a new gauge group called ``hypercolor'' as described in~\cite{Kilic:2008ub,PhenoOctet,Bai:2011mr}, 
and 
extended supersymmetric models~\cite{Choi:2008ub,Kribs:2007ac,Plehn:2008ae}. 

In the hypercolor model, new colored fermions are \linebreak charged under an additional gauge group (SU(3) hypercolor).
In analogy with QCD, new colored fermions are bound into mesons of hypercolor: a color octet vector particle, the coloron, and a color octet pseudoscalar, 
the hyperpion which decays to two gluons. 

In supersymmetric models with Dirac gluinos, a scalar gluon (sgluon) extends the QCD sector, which is made
up of the gluon/gluino super-multiplet and an additional gluino/sgluon super-multiplet. As the sgluon has positive R-parity~\cite{Farrar:1978xj},
light sgluons, i.e. sgluons with masses of the 
order of 100~\GeV, are expected to decay to two gluons with a branching ratio close to 1.
Single production of sgluons, loop-induced via supersymmetric particles, is also possible, 
but these cross sections are several orders of magnitude smaller than those of the studies in Ref.~\cite{Collaboration:2011fq}; therefore the previously obtained limit
of 1.92~\TeV\ on the color-octet mass does not apply to the models studied in this letter.
The pair production cross section 
does not depend, at leading order, on supersymmetric parameters
except the sgluon mass.

In the following, the sgluon pair production will be used as the benchmark process for the production of two heavy objects of equal mass, each decaying into two jets.

\mbox{ATLAS}~\cite{atlas_detector} is a multipurpose detector with nearly $4\pi$ coverage in solid angle.
The inner detector, consisting of 
silicon pixel and microstrip detectors as well as a transition radiation tracker,
is immersed in a 2~T solenoidal magnetic field. 
The finely-segmented, hermetic calorimeter  
covers $|\eta| < 4.9$\footnote{ATLAS uses a right-handed coordinate system with its origin at the nominal interaction point (IP) in the centre of the detector and the $z$-axis along the beam pipe. The $x$-axis points from the IP to the centre of the LHC ring, and the $y$ axis points upward. Cylindrical coordinates $(r,\phi)$ are used in the transverse plane, $\phi$ being the azimuthal angle around the beam pipe. The pseudorapidity is defined in terms of the polar angle $\theta$ as $\eta=-\ln\tan(\theta/2)$.}
 and provides three-dimensional reconstruction 
of particle showers.  
The electromagnetic calo\-rimeter is 
a lead liquid-argon sampling calorimeter. In the central region it is 
surrounded by a hadronic calorimeter made of iron and scintillating tiles. 
For the region $|\eta| > 1.7$ the sampling calorimeter consists of copper or tungsten and
liquid argon. 
The calorimeters are surrounded by the muon spectrometer
which consists
of three large superconducting toroids, a system of
precision tracking chambers, and detectors for triggering.

ATLAS uses a three-level trigger system. The first level trigger is implemented in hardware, the other two trigger
levels are implemented in software. In the analysis particular emphasis was placed on being sensitive to the low mass region
in order to exploit the low trigger thresholds of the data recorded in 2010. 
Therefore, a first level trigger requiring at least four 
jets with a transverse momentum $p_\mathrm{T}$ of at least 5~\GeV\ was used. 
The trigger efficiency increases as a function of the jet $p_\mathrm{T}$ to at least 99\%~\cite{ATLAS-CONF-2010-094} for calibrated jets of $p_T>$ 55~\GeV.
An integrated luminosity of 34~\ipb\ has been recorded with this trigger, taking into account the prescaling of the trigger at the end of 2010. 

The Standard Model (SM) multijet production of four jets with $p_T>60$~\GeV\ and $|\eta|<2.8$ has a cross section of 
approximately 5~nb~\cite{Collaboration:2011tq}. 
Compared to multijet production, even without considering the branching ratios to obtain
a final state with four hard jets, 
other Standard Model backgrounds have much smaller cross sections: \Wboson\Wboson\ with a cross section of 41~pb~\cite{Aad:1347441}, \ttbar\ with a cross section of 171~pb~\cite{Aad:2011yb} and \Wboson~production associated with two jets 
($p_\mathrm{T}> $20~\GeV, $\eta<$2.8) with a cross section of 200~pb~\cite{Aad:1318461}. 

Monte Carlo (MC) samples were used to model the SM multijet background.
ALPGEN~\cite{Mangano:2002ea} SM multijet samples were generated with the MLM matching scheme~\cite{Alwall:2007fs}, interfaced to HERWIG~\cite{Corcella:2002jc} for parton shower and
fragmentation and to JIMMY~\cite{Butterworth:1996zw} for the simulation of the underlying event. 
As a cross check, PYTHIA~\cite{pythia} SM dijet samples were generated with the MRST~LO$^*$ Parton Density Function (PDF)~\cite{Sherstnev:2007nd} and the ALPGEN sample with CTEQ6L1 PDF~\cite{Pumplin:2002vw}.
For both, the underlying event tune was the ATLAS MC09 tune~\cite{ATLAS-PHYS-PUB-2010-002}.
The sgluon pair production differential 
cross section of Ref.~\cite{Choi:2008ub} was implemented as an external process to PYTHIA. 
Signal samples of 10k events each for M$_{\mathrm{sgluon}}$=100 to 200~\GeV\ in steps of 10~\GeV\ and a point at 225~\GeV\
were generated.
The cross section for the pair production of sgluons, computed with the MRST~LO$^*$ PDF, is 7.5~nb at 100~\GeV. 
It decreases to 100~pb at 225~\GeV.
All signal and SM background samples were passed through a {\sc GEANT4}~\cite{Agostinelli:2002hh} based simulation of the ATLAS detector.

\indent Collision candidates are selected by requiring at least one reconstructed proton-proton interaction vertex with more than four good quality tracks.
Jets are reconstructed using the anti-k$_t$ jet clustering algorithm~\cite{cacciari-2008} with a radius parameter of 0.6. The inputs to the jet algorithm are three-dimensional clusters formed from energy deposits in the calorimeter. The jets are calibrated using $p_\mathrm{T}$- and $\mathrm{\eta}$-dependent correction factors based on MC simulation and validated by test beam and collision data studies~\cite{ATLAS-CONF-2011-032}.
All jets considered in the analysis must have $p_\mathrm{T}>20$~\GeV~and $|\eta|<2.8$. 
Quality selections are applied to the reconstructed jets to eliminate various detector effects and suppress beam and other non-collision backgrounds. 
Overall, these requirements reduce the sample size by less than 0.1\%. 

At least four jets are required with a $p_\mathrm{T}$ greater than 55\% of the sgluon mass 
to improve the background rejection. 
The analysis thus probes sgluon masses greater than 100~\GeV\ due to the trigger.
After this requirement, as the system has a non-negligible boost leading to a jet-jet separation of $\Delta R_{jj} \approx 1$ 
(where $\Delta R_{jj}=\sqrt{(\Delta \phi_{jj})^2+(\Delta \eta_{jj})^2} $), 
the four highest $p_\mathrm{T}$ jets in the event are paired by minimizing $|\Delta R_{\mathrm{pair1}} -1|  +| \Delta R_{\mathrm{pair2}}  - 1|$. 
Events are rejected if, for the chosen combination, a jet-jet-pairing has $\Delta R_{jj}>1.6$. 
The corresponding reconstructed masses are denoted $M_1$ and $M_2$ in the following and the 
reconstructed average mass is defined as $((M_1+M_2)/2)$.
Finally, to improve further the rejection of the SM multijet background, the relative difference between the
two reconstructed masses ($|M_1 - M_2|/(M_1+M_2)$) is required to be less than 7.5\%. 
The scattering angle ($|\cos(\theta^*)|$) of the reconstructed sgluons in the rest frame of the four leading jets is required to be less than 0.5.
The SM multijet background is peaked in the forward region, reflecting t-channel gluon exchange, while the signal is 
produced centrally due to the scalar nature of the sgluon.
The selection efficiency of the signal is 0.62\% for a sgluon mass of 100~\GeV\ and 0.32\% for a mass of 225~\GeV.

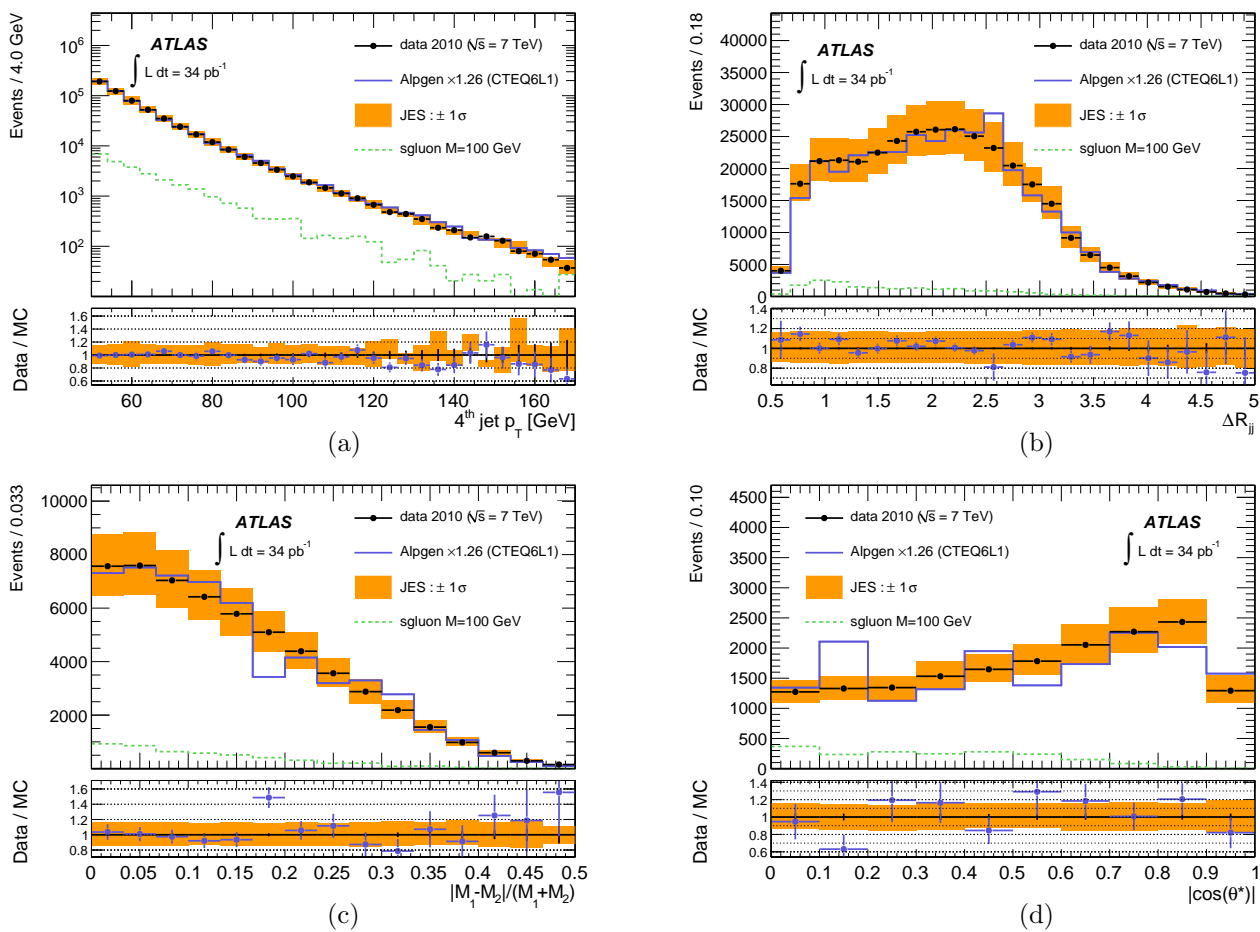
\begin{figure*}[htb]
   \begin{minipage}[c]{.49\linewidth}
      \includegraphics[scale=0.43]{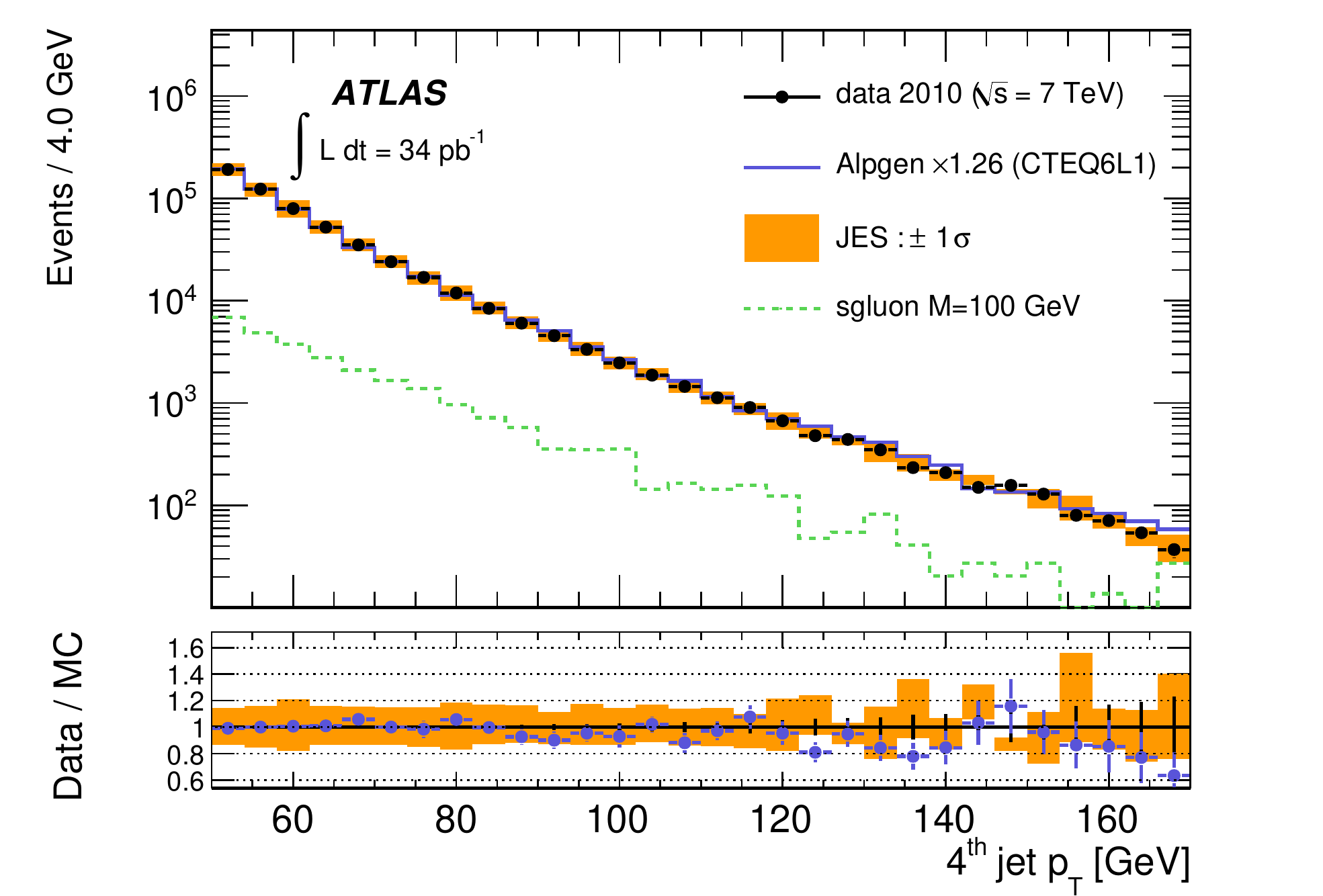}
   \end{minipage} \hfill
   \begin{minipage}[c]{.49\linewidth}
      \includegraphics[scale=0.43]{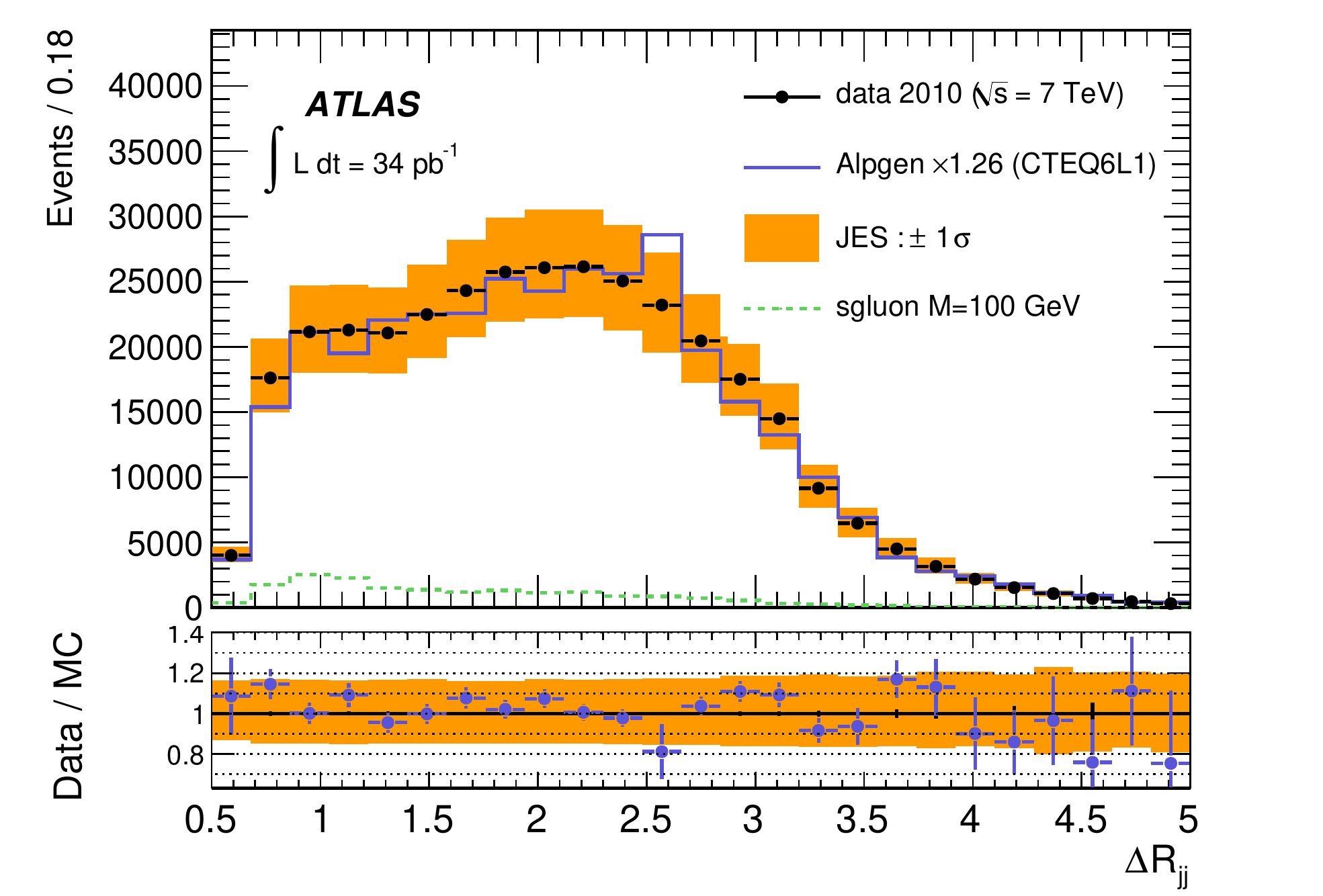}
   \end{minipage}
   \begin{picture} (0.,0.)
		\setlength{\unitlength}{1.0cm}
		\put (-13.4, -3){(a)}
		\put (-4.3,  -3){(b)}
   \end{picture}
    \vspace*{0.2cm}
    \vspace*{0.2cm}
   \begin{minipage}[c]{.49\linewidth}
      \includegraphics[scale=0.43]{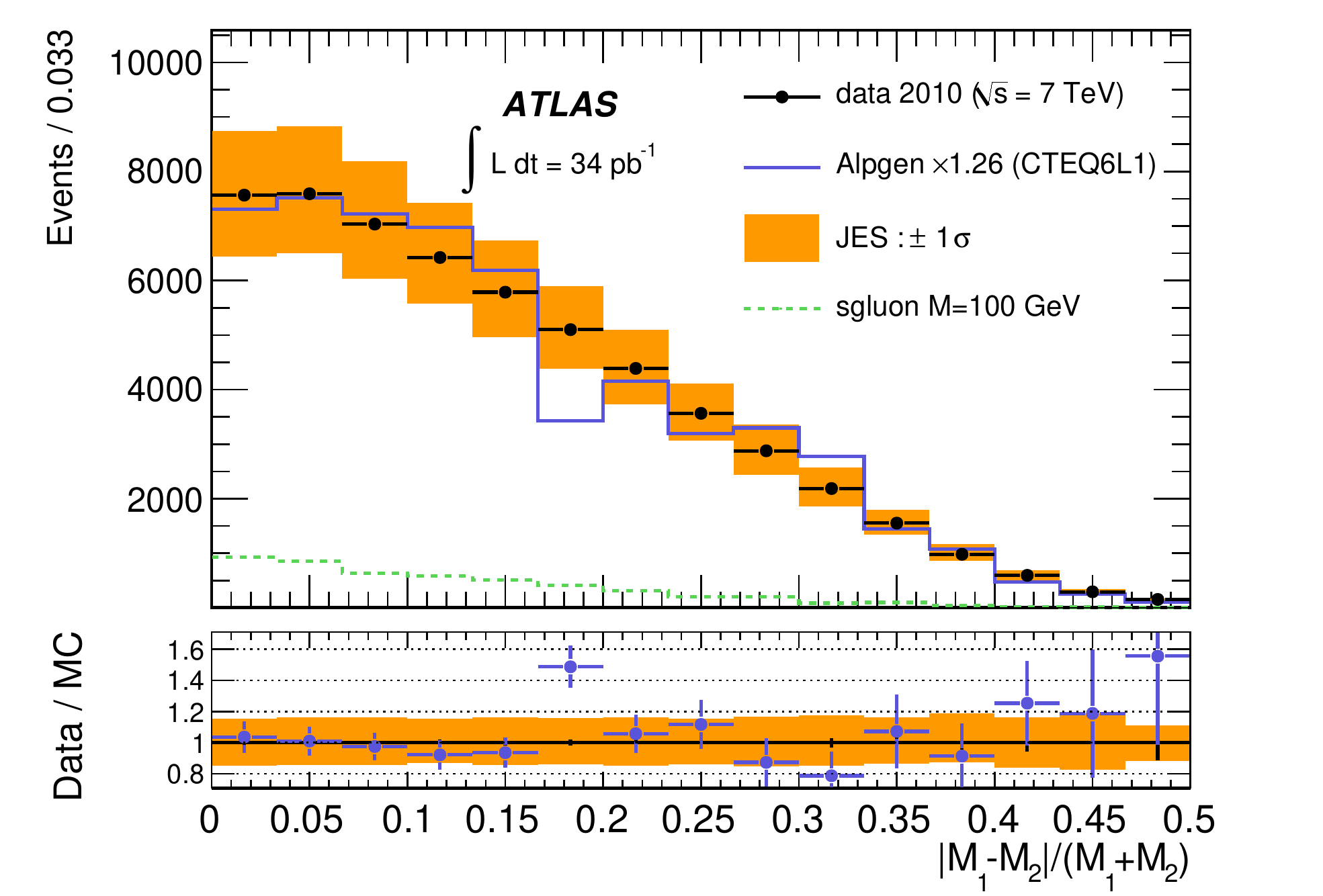}
   \end{minipage} \hfill
   \begin{minipage}[c]{.49\linewidth}
      \includegraphics[scale=0.43]{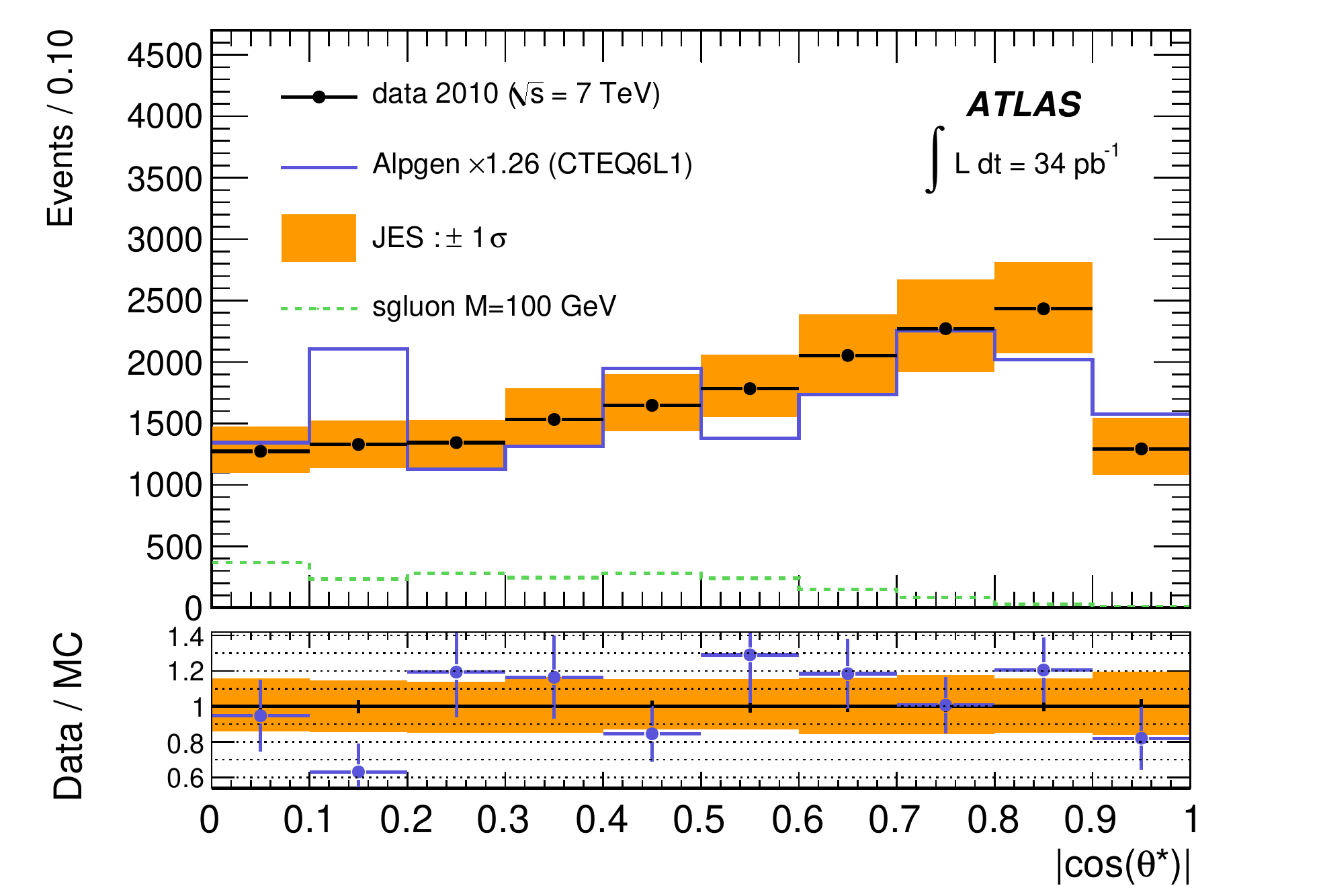}
   \end{minipage}
   \begin{picture} (0.,0.)
		\setlength{\unitlength}{1.0cm}
		\put (-13.4, -3){(c)}
		\put (-4.3,  -3){(d)}
   \end{picture}
    \vspace*{0.2cm}
    \vspace*{0.2cm}

   \caption{\small Kinematic variables at different stages of the analysis. Data (dots) are compared to the ALPGEN MC sample (solid line). 
The solid band corresponds to one standard deviation in the jet energy scale. The ratio data/MC is also shown with its statistical uncertainty, which is dominated by the MC statistics.
The dashed line corresponds to a sgluon signal of 100~\GeV. 
(a) The transverse momentum of the 4$^{th}$ jet is shown. 
(b) The $\Delta R_{jj}$ distribution for the reconstructed sgluon candidate with the highest transverse momentum jet is shown after requiring the transverse momentum to be greater than 55~\GeV\ and pairing the four leading jets into two sgluon candidates. 
(c) The relative mass difference is shown after the criterias on the $p_\mathrm{T}$ and $\Delta R_{jj}$ have been applied. 
(d) The scattering angle in the 4-jet center-of-mass frame is shown after all other selection criteria have been applied.}
   \label{fig:data_mc}
\end{figure*}

The variables used in the analysis are compared to the ALPGEN MC sample (solid line) 
at different stages of the analysis in Fig.~\ref{fig:data_mc}, as well as to the simulated signal
sample for a sgluon of mass~100~\GeV\ (dashed line). 
The data to MC ratios are shown below each figure. The solid band corresponds to one standard deviation in the jet energy scale (JES)~\cite{ATLAS-CONF-2011-032}. 
Figure~\ref{fig:data_mc} (a) shows the $p_\mathrm{T}$ of 
the $4^{th}$ leading jet ordered by $p_\mathrm{T}$ for events with 4~jets with $p_\mathrm{T}>50$~\GeV. 
The ALPGEN MC sample is normalized to the data after this requirement (normalization factor~1.26). The same normalization is used for all the following figures. This factor is similar to the one obtained in~\cite{Collaboration:2011tq} for a different event selection.
Figure~\ref{fig:data_mc} (b) shows the $\Delta R_{jj}$ distribution between the two jets from the sgluon reconstructed with the highest $p_\mathrm{T}$ jet in the event. 
The $4^{th}$ leading jet was required to have a $p_\mathrm{T}$ greater than 55~\GeV. 
The relative mass difference is shown in Fig.~\ref{fig:data_mc} (c) after the application of the selection criteria on the $p_\mathrm{T}$ and the $\Delta R_{jj}$. The scattering angle is shown in Fig.~\ref{fig:data_mc} (d) after all other criteria are applied.

\begin{table*}[htb]
\caption{\small Cut flow for data, ALPGEN MC sample and sgluon MC (M$_{\mathrm{sgluon}}=100$~\GeV). The ALPGEN MC sample is normalized to the data after the first requirement.}
\label{tab:cutflow}
\centering
\begin{tabular}{lrcc|c}
\hline
\hline
Cut & data & ALPGEN MC & sgluon MC & sgluon/ALPGEN\\
\hline
4 jets p$_T>50$~\GeV          & 568421   & 568000 $\pm$ 8000 & 27900 $\pm$ 800 & 4.9\% \\
4 jets p$_T>55$~\GeV          & 340429   & 336000 $\pm$ 6000 & 19000 $\pm$ 700 & 5.6\% \\
$\Delta R_{jj}<1.6$     & 56131    & 55400 $\pm$ 1900  & 4900 $\pm$ 350  & 8.8\%    \\ 
$|M_1 - M_2|/(M_1+M_2)<0.075$ & 16958    & 16800 $\pm$ 1100  & 1910 $\pm$ 220  & 11.4\% \\
$|\cos(\theta^*)|<0.5$         & 6937    & 7700 $\pm$ 800    & 1450 $\pm$ 190  & 18.9\% \\
\hline
\hline
\end{tabular}
\end{table*}

The shape of the data is well described by the ALPGEN MC simulations. As the jet energy scale error 
can be considered to be, at first order, an error on the normalization, it is important to note
that the description is good even when considering only the statistical errors 
in Fig.~\ref{fig:data_mc}. The cut flow is shown in 
Table~\ref{tab:cutflow} for the data, the ALPGEN MC sample and a sgluon mass of 100~\GeV. 

While the description of the data by the MC is satisfactory, but limited by the MC statistics, 
the background is derived from data alone by dividing the data sample into one signal
region (A) and three background-dominated regions (B, C, D). The variables used to define the different regions 
are the sgluon scattering angle and the relative mass difference. 
The regions, defined in Table~\ref{tab:defabcd}, are chosen as a compromise between low contamination from the signal
and the statistical error in the regions~B, C and D which feeds into the error on the background prediction. 

For the average mass distribution, the shape of the background in the signal region is modeled by that in the control region B, parameterized by a fit to
$f(x) = (x - p_1)^{p_2} \cdot e^{- x \cdot p_3 -x^2 \cdot p_4}$ where
$p_1$, $p_2$, $p_3$, $p_4$ are the free parameters and $x$ is the reconstructed average mass. 
The parameters $p_1$ and $p_2$ describe the rising edge of the distribution; whereas, the $p_3$ and $p_4$ parameters model its tail.
In region~B the selection criteria on the scattering angle is inverted with respect to the signal region, but not the one on the relative mass difference, leading to the best description of the background shape in the signal region.

The normalization of the background in the signal region is derived from the number of events in the control regions: $N_A^{\mathrm{extrapolation}}=N_B\cdot N_C/N_D$. This is referred to as the ABCD method.
The effect of the correlation between the two variables used to define the different regions is neglected since the correlation is less than 0.2\%. Performing a closure test on the ALPGEN and PYTHIA MC samples, the number of events
in the signal region agreed with the prediction derived from the three other regions within the statistical error. 

\begin{table}[htb]
\caption{\small Definition of the four regions for the background determination. Region A is the signal region.}
\centering
\small
\begin{tabular}{cc}
\hline
\hline
Region & Selection \\
\hline
A & $|\cos(\theta^*)|<0.5$ and $|M_1 - M_2|/(M_1+M_2)<$7.5\%\\
B & $|\cos(\theta^*)|>0.7$ and $|M_1 - M_2|/(M_1+M_2)<$7.5\% \\
C & $|\cos(\theta^*)|<0.5$ and $|M_1 - M_2|/(M_1+M_2)>$7.5\% \\
D & $|\cos(\theta^*)|>0.7$ and $|M_1 - M_2|/(M_1+M_2)>$7.5\% \\
\hline
\hline
\end{tabular}
\label{tab:defabcd}
\end{table}

\begin{figure*}[htb]
   \begin{minipage}[c]{.49\linewidth}
      \includegraphics[scale=0.43]{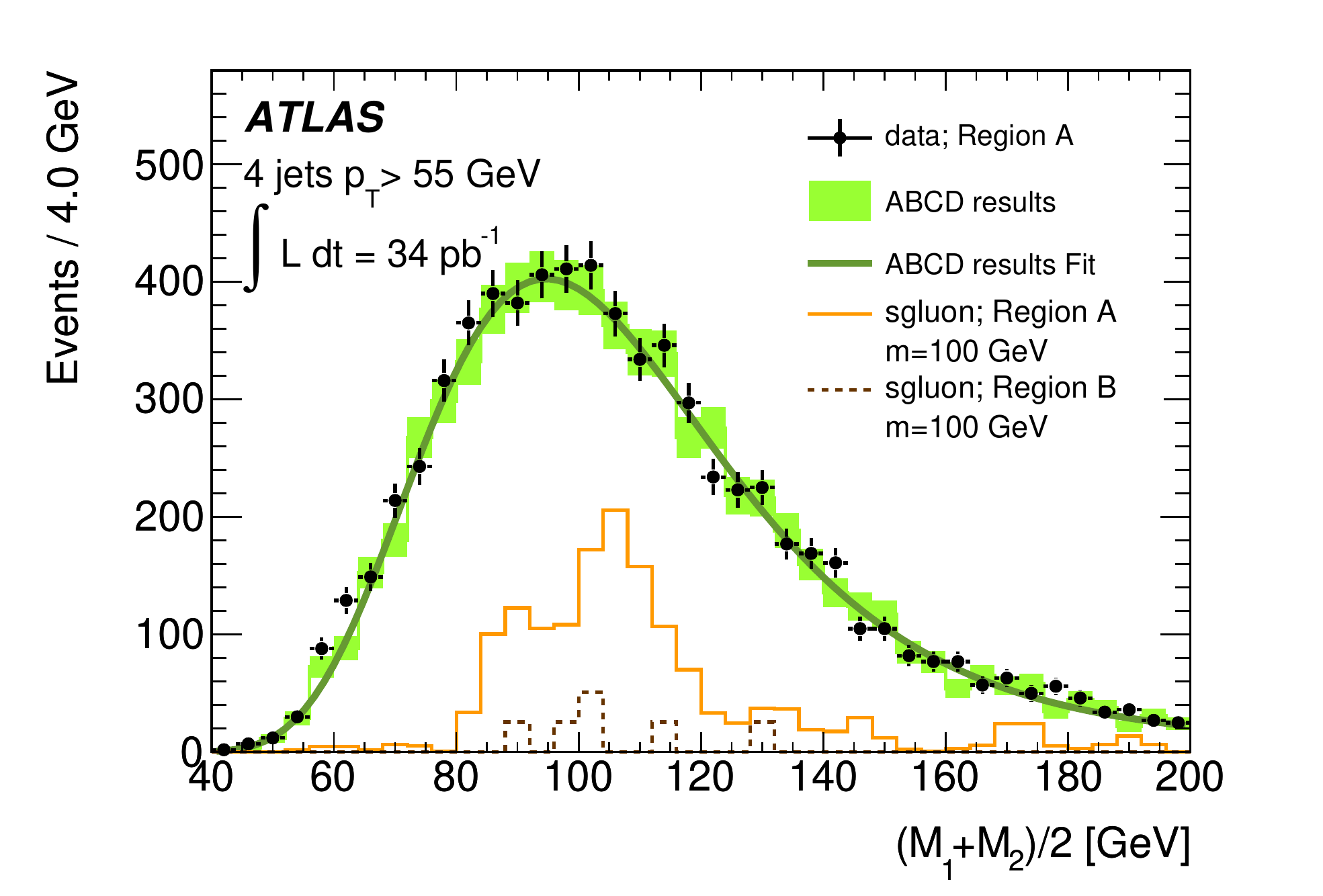}
   \end{minipage}\hfill
      \begin{minipage}[c]{.49\linewidth}
   	  \includegraphics[scale=0.43]{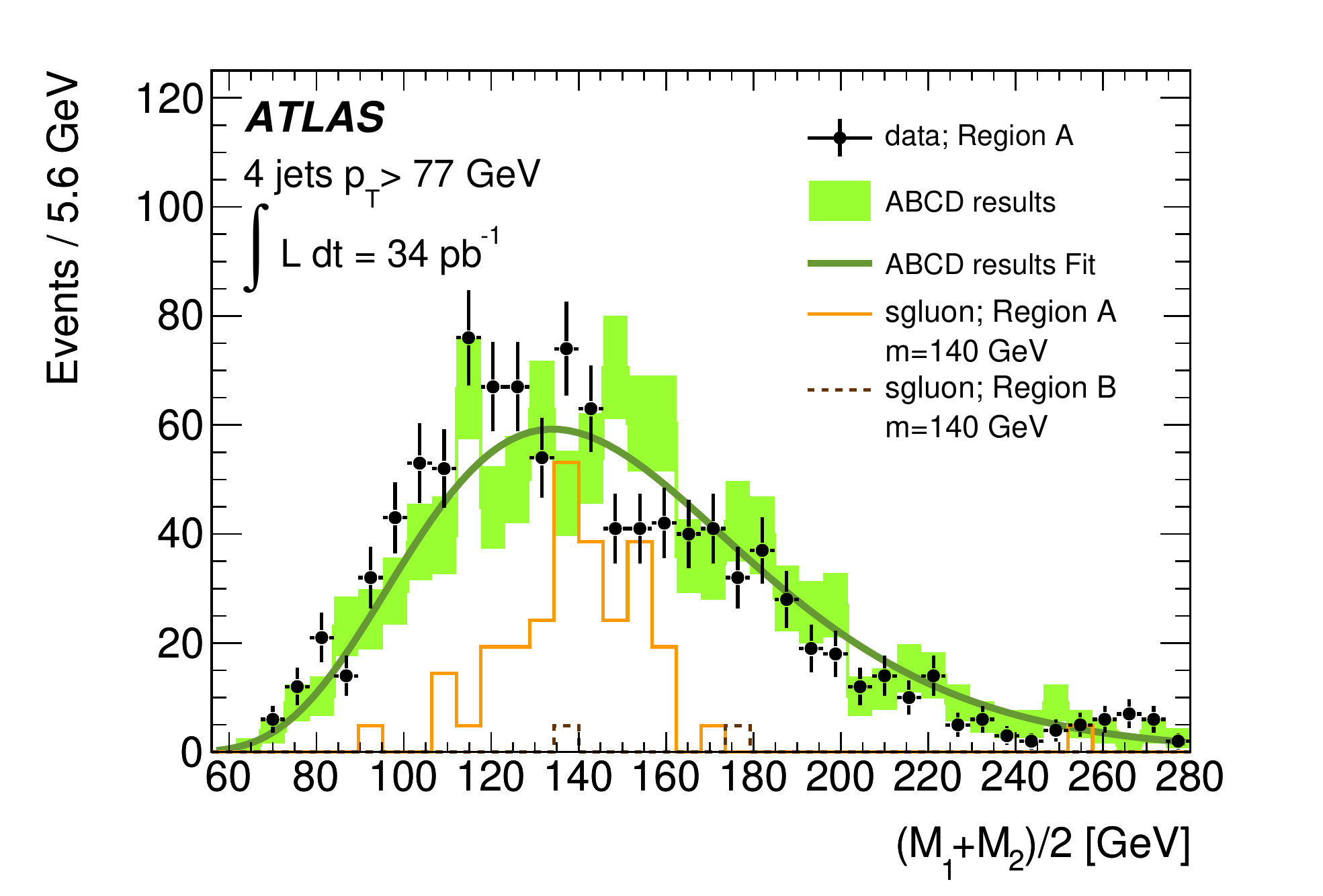}
   \end{minipage} 
   \begin{picture} (0.,0.)
		\setlength{\unitlength}{1.0cm}
		\put (-13.4, -3){(a)}
		\put (-4.3,  -3){(b)}
   \end{picture}
    \vspace*{0.2cm}
    \vspace*{0.2cm}
   \begin{minipage}[c]{.49\linewidth}
      \includegraphics[scale=0.43]{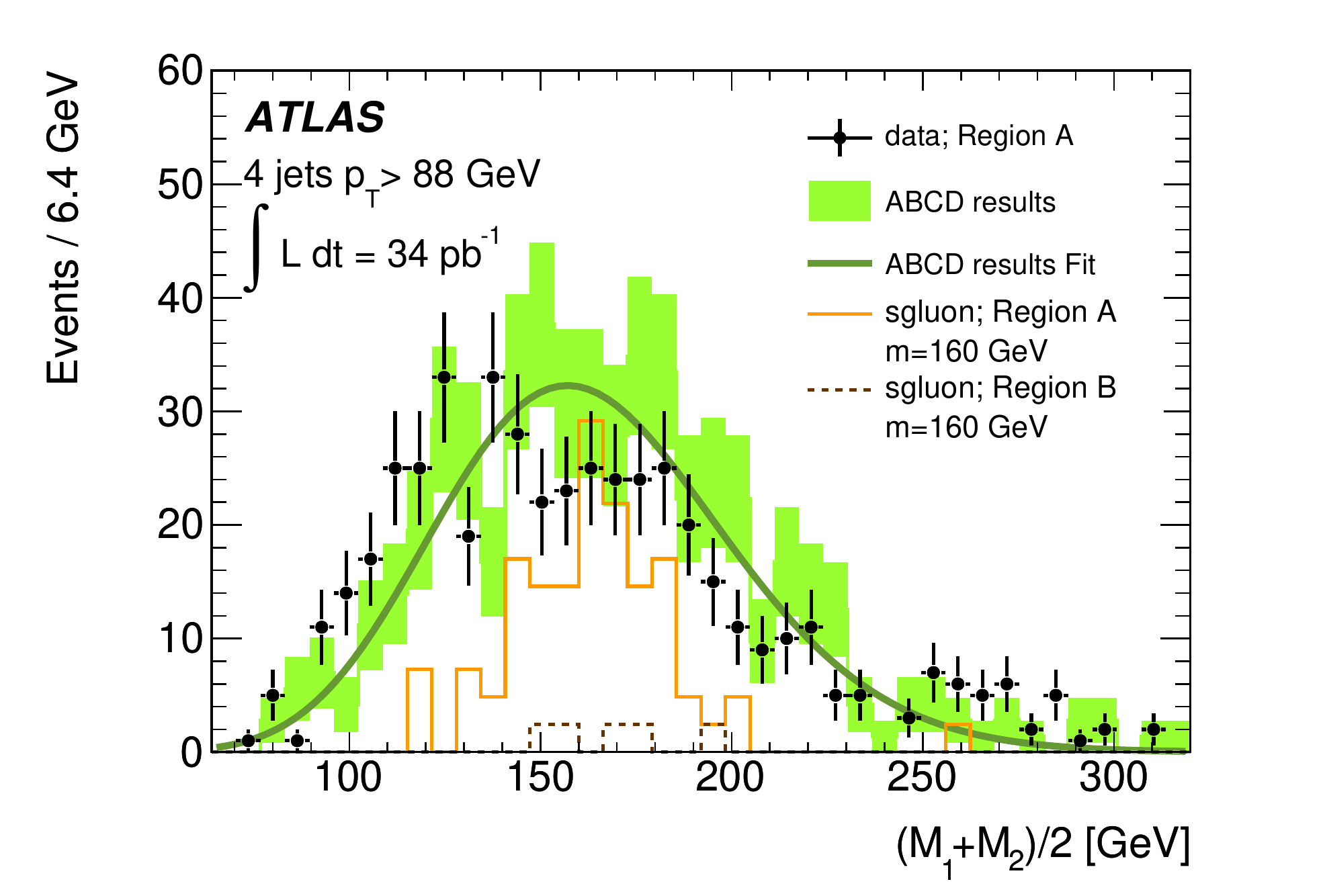}
   \end{minipage}\hfill
   \begin{minipage}[c]{.49\linewidth}
      \includegraphics[scale=0.43]{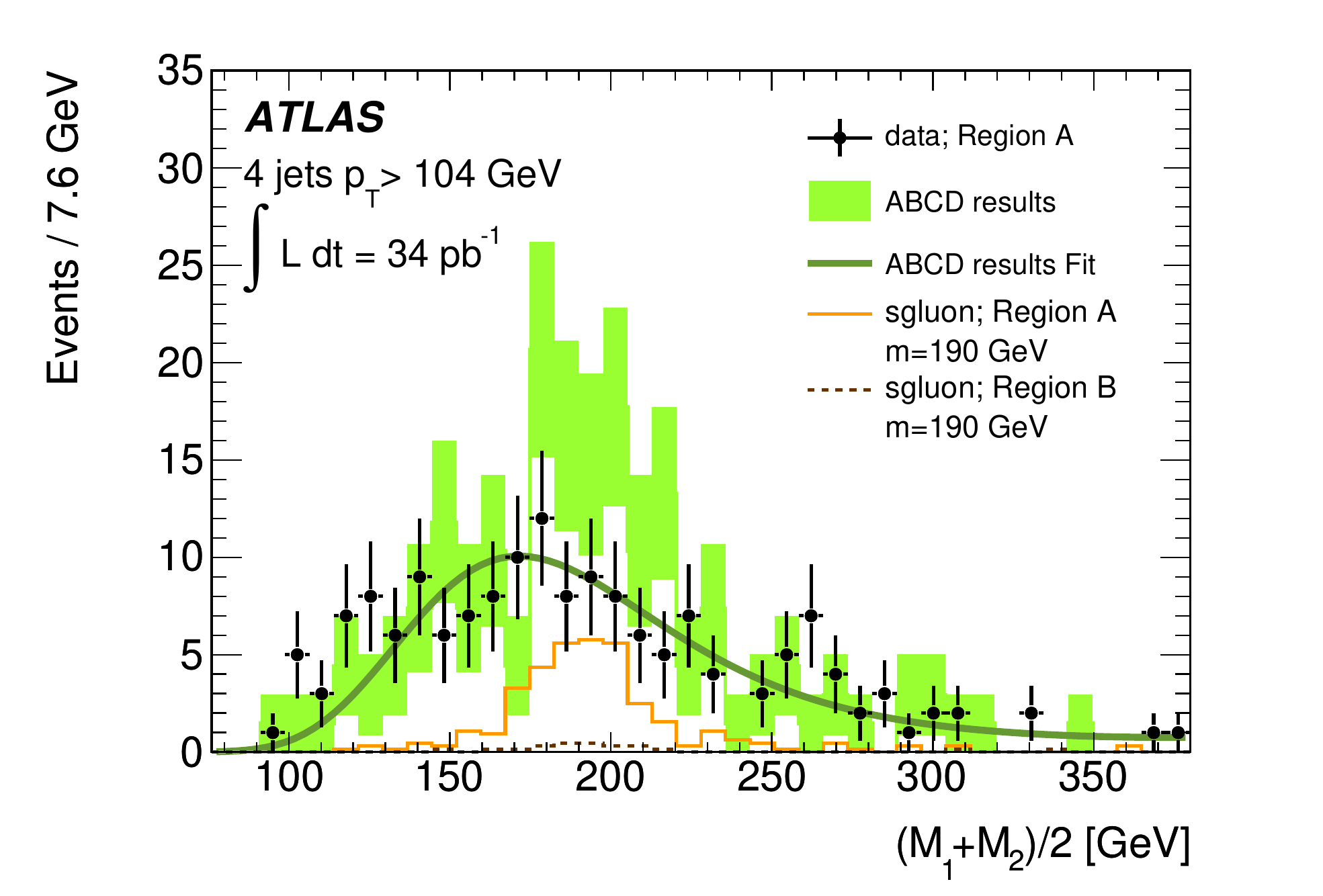}
   \end{minipage}
   \begin{picture} (0.,0.)
		\setlength{\unitlength}{1.0cm}
		\put (-13.4, -3){(c)}
		\put (-4.3,  -3){(d)}
   \end{picture}
    \vspace*{0.2cm}
    \vspace*{0.2cm}
  \caption{The comparison of the prediction of the background with the data in the signal region is shown. The points are the data in the signal region (region A). The solid (dashed) histogram is the estimated signal in region~A~(B) for the nominal cross section. The predictions of background in region A based upon the data in region B (rectangles) and upon the result of the fit in region B (line), each normalized using the ABCD method, are shown for: (a) $M_{\mathrm{sgluon}}=100$~\GeV, (b) $M_{\mathrm{sgluon}}=140$~\GeV, (c) $M_{\mathrm{sgluon}}=160$~\GeV\ and  (d) $M_{\mathrm{sgluon}}=190$~\GeV. For each, the bin size of the histogram is equal to $0.04\times M_{\mathrm{sgluon}}$.}
  \label{fig:res_matrix}
\end{figure*}
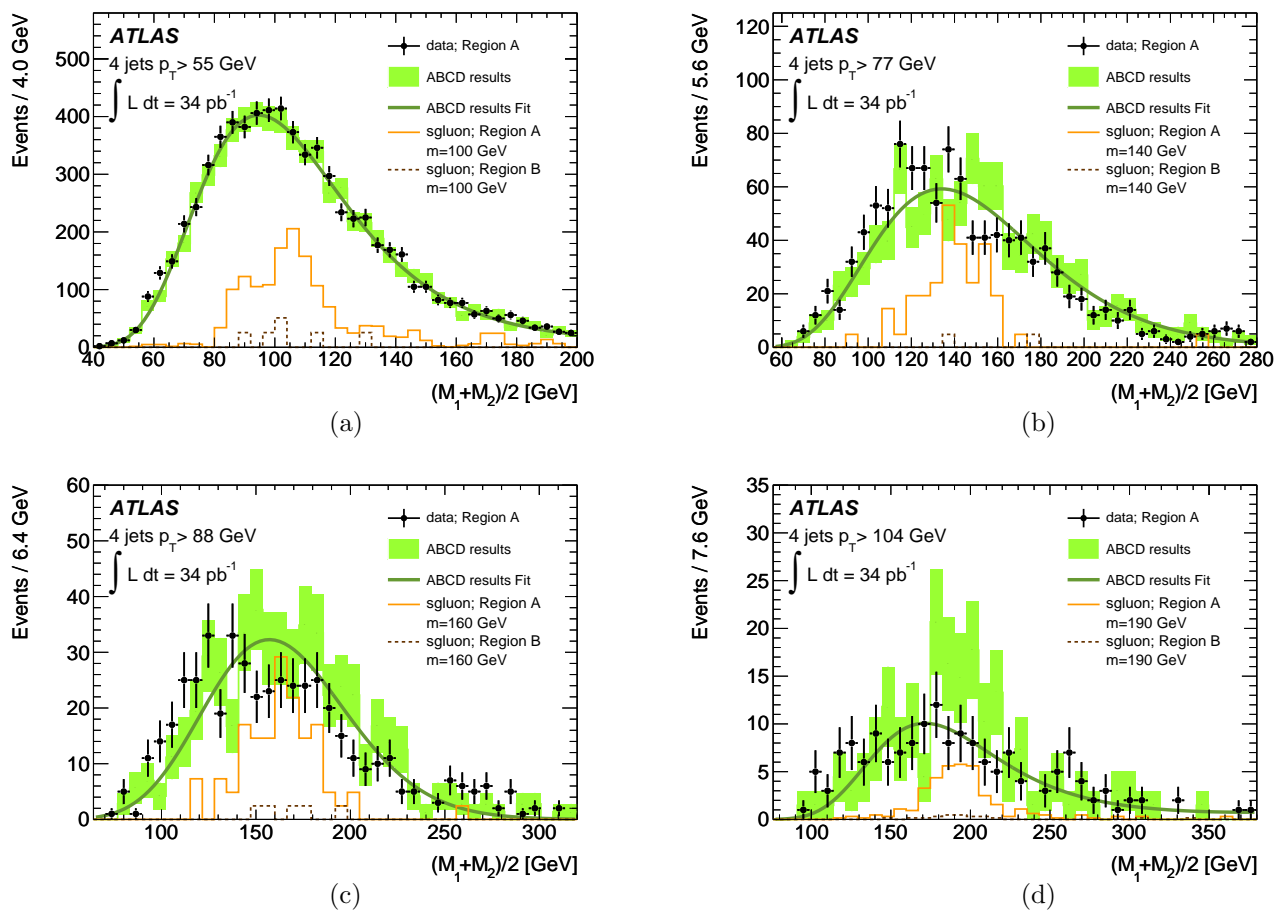

\begin{table*}[htb]
\caption{\small Comparison of data in signal region with background prediction. The first column shows the $p_\mathrm{T}$ requirement applied on the $4^{th}$ leading jet in $p_\mathrm{T}$, the second column the observed number of events in the signal region. 
The third column shows the prediction of the ABCD method. Only the statistical uncertainty is indicated. 
The fourth column is the $\chi^{2}/NDF(A,B)$ between the shapes of the reconstructed average mass distribution in regions A and B. The last column shows $\chi^2/NDF(B)$ 
for the fit to the background region.}
\centering
\begin{tabular}{crccc}
\hline
\hline
$p_T^{\rm min}$ (4$^{th}$ jet) [GeV] & data & ABCD prediction & $\chi^2/NDF(A,B)$  & $\chi^2/NDF(B)$\\
\hline
 49    &    $11732$  &    $ 11410 \pm 150 $  &   1.31   & 0.77 \\
 55    &    $6937$   &    $ 6740 \pm 120 $   &   1.02   & 1.05 \\
 60    &    $4098$   &    $ 3980 \pm 90 $    &   0.85   & 1.09 \\
 66    &    $2532$   &    $ 2460 \pm 70 $    &   1.04   & 0.87 \\
 71    &    $1590$   &    $ 1580 \pm 60 $    &   1.18   & 0.98 \\
 77    &    $1069$   &    $ 1030 \pm 50 $    &   1.39   & 0.61 \\
 82    &    $701$    &    $ 720 \pm 40 $     &   1.59   & 1.04 \\
 88    &    $480$    &    $ 517 \pm 34 $     &   1.32   & 1.00 \\
 93    &    $322$    &    $ 364 \pm 29 $     &   0.94   & 1.22 \\
 99    &    $218$    &    $ 266 \pm 25 $     &   1.08   & 1.22 \\
 104   &    $162$    &    $ 187 \pm 21 $     &   1.05   & 1.13 \\
 110   &    $116$    &    $ 151 \pm 19 $     &   1.42   & 1.44 \\
\hline
\hline
\end{tabular}
\label{tab:matres}
\end{table*}

In Fig.~\ref{fig:res_matrix}  the result obtained with the data is shown in the signal region (region A)
for sgluon masses of 100, 140, 160 and 190~\GeV.
The data in region A is compared to the data in the control region B, and to the fit in region B, where the data and the fit are each normalized using the ABCD method.
The expected signals in regions A and B, normalized to the nominal sgluon cross section, are also shown. 
Table~\ref{tab:matres} shows the number of events in the signal region, the prediction of the background
from the ABCD prediction, the $\chi^{2}$ per degree of freedom ($NDF$) 
between the shapes of the distributions in region A and B~($\chi^{2}/NDF(A,B)$), as well as the $\chi^{2}/NDF(B)$ in the background region for the fit of the background function.
No significant deviation is observed between the data-driven background prediction and the data.
Therefore limits are set on the excluded cross section using a profile likelihood ratio with the CL$_s$ approach~\cite{Read:2002hq}.
The shapes of the average mass distribution for signal and background in region~A are parametrised and used in the likelihood. 
The signal shape is modeled with a Gaussian distribution and the background shape with the parametrisation of the ABCD method. The signal contamination in the control regions is taken into account in the likelihood. 
A Gaussian distribution is used to simulate the signal contamination in region~B; whereas, in region~C and~D no assumption is made on the signal shape since only the number of events is used in these two regions.

\begin{table*}[htb]
\caption{\small The systematic uncertainties due to the jet energy scale (JES), jet energy resolution (JER),
the ABCD method (ABCD), the choice of the PDF (PDF), the integrated luminosity (L), the Monte Carlo statistics (MC stat.)
and the trigger efficiency (Trigger).}
\centering
\begin{tabular}{l|c}
\hline
\hline
Source & Effect \\
\hline
JES   & Signal peak center $\pm 2\% $ \\
      & Signal efficiency  $\pm 15\% $      \\
\hline
JER & Signal peak width $\pm 10 \%$ \\
\hline 
ABCD & Background prediction $\pm 1\%$  to $\pm 10\% $\\
\hline
PDF & Signal efficiency  $\pm 2\%$    \\
\hline
L & Signal normalization $\pm 3.4\%$ \\
\hline
MC stat. &  Signal normalization in A(B,C,D) $\pm 5(16,5,16) \%$ \\
\hline
Trigger & Signal normalization  ($\mathrm{eff} = 99$\%) $\pm 1\%$ \\
\hline
\hline
\end{tabular}
\label{tab:sys}
\end{table*}

The different sources of systematic uncertainty and their effect are summarized 
in Table~\ref{tab:sys}. 
The uncertainty on the integrated luminosity is 3.4\%~\cite{Aad:2011dr}. The trigger efficiency
is estimated in minimum bias data 
to be 99$\pm$1\%. The signal acceptance and contamination are taken from the full simulation Monte Carlo samples 
with a statistical uncertainty of 5\% (in region~A) by fitting the efficiencies as a function of the sgluon mass. 
The jet energy scale uncertainty is propagated to 
the signal~\cite{ATLAS-CONF-2011-032}, affecting the selection efficiency by 15\%. 
A second effect of the JES uncertainty on the signal is a $\pm 2\%$ shift of the signal mass peak position.  
The impact of the jet energy resolution uncertainty on the signal mass peak width is 10\%. 
The impact of the choice of the PDF for the signal generation was estimated to be less than 2\%. Finally a systematic error, reflecting the statistics available to check the prediction of the ABCD method in the absence of new physics, is assigned to the background prediction.
Gaussian nuisance parameters are implemented in the likelihood  
corresponding to the errors taking into account the correlations, e.g. 
the error on the luminosity is common to the ABCD regions. 
The contamination of the regions B, C and D by the signal is also taken into account in the likelihood.

\begin{figure}[htb]
  \centering
  \includegraphics[scale=0.43]{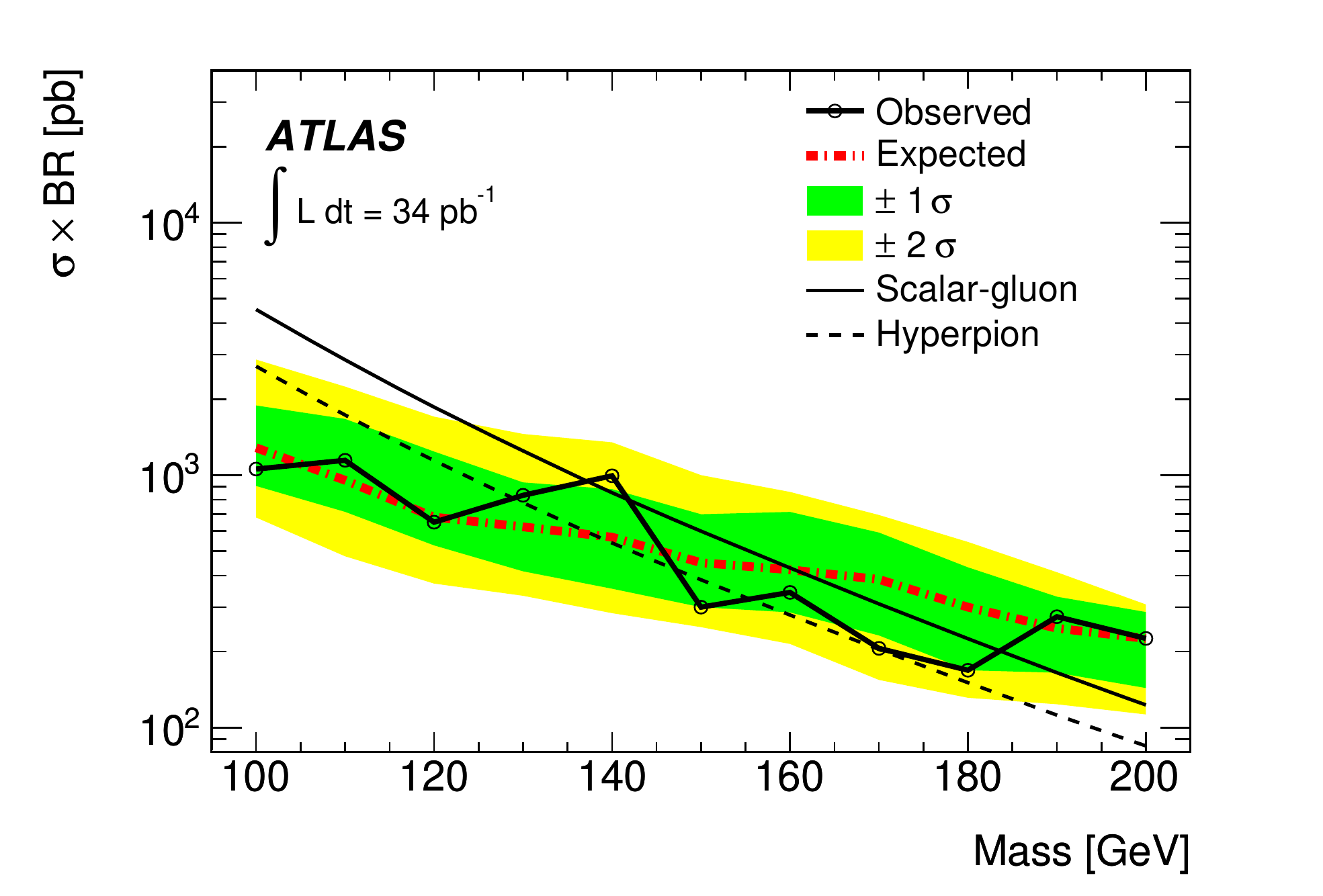}
  \caption{\small Expected and observed 95\% CL upper bounds on the product of the scalar pair production cross sections and of the branching ratio to gluons as a function of the scalar mass. The predictions of the sgluon and hyperpion pair production cross section are also shown.}
  \label{fig:exclu}
\end{figure}

For each tested mass, the observed and expected median CL$_s$ are determined as a function of the signal cross section. 
The analysis is performed for masses from 100 to 200~\GeV~in steps of 10~\GeV. 
The resulting excluded cross section, shown in Fig.~\ref{fig:exclu}, is 1~nb at 100~\GeV~and 280~pb at 190~\GeV. 
Converting this result into a mass limit for a branching ratio of 1 to gluon pairs, using a leading order cross section~\cite{Choi:2008ub} with CTEQ6L1~\cite{Pumplin:2002vw}, sgluons with masses from 100~\GeV\ to 185~\GeV\ are excluded at 95\% confidence level with the exception of a mass window of about 5~\GeV\ around 140~\GeV. 
The sgluon cross section used was checked at $\sqrt{s}=14$~\TeV\ with Ref.~\cite{Choi:2008ub} and was found to agree at the percent level.
The centrality of the hyperpions compared to the sgluons increases due to the additional contribution of the s-channel coloron exchange. 
This property should increase the selection efficiency due to the presence of the requirement on the scattering angle.
However, the hyperpion cross section was scaled down from the sgluon cross section according to Ref.~\cite{PhenoOctet}, which makes the limits less stringent.
Hyperpions with masses of 100~\GeV\ to 155~\GeV\ are excluded with the exception of a mass window
of 15~\GeV\ around 140~\GeV.

\indent To conclude, four-jet events have been analyzed by the ATLAS experiment, searching for the pair production of a 
new scalar particle decaying to two jets. The data in the signal region is in good agreement with the data-driven background estimation. 
No evidence for new phenomena was found. 
Cross section limits as a function of the mass of the scalar particle have been determined.
Interpreting the cross section limit, sgluons (hyperpions) 
with masses from 100~\GeV\ to 185~\GeV\ (155~\GeV) are excluded at 95\% CL.
A mass window of about 5~\GeV\ (15~\GeV) around 140~\GeV\ remains unexcluded for the sgluons (hyperpions). 

\indent We would like to thank the authors of Refs.~\cite{Kilic:2008ub,Choi:2008ub}, 
in particular S. Schumann and P.M. Zerwas, for useful discussions. We are grateful to E.~Popenda
for having provided us with the code for the differential cross section of sgluon pair production.

We thank CERN for the very successful operation of the LHC, as well as the
support staff from our institutions without whom ATLAS could not be
operated efficiently.

We acknowledge the support of ANPCyT, Argentina; YerPhI, Armenia; ARC,
Australia; BMWF, Austria; ANAS, Azerbaijan; SSTC, Belarus; CNPq and FAPESP,
Brazil; NSERC, NRC and CFI, Canada; CERN; CONICYT, Chile; CAS, MOST and
NSFC, China; COLCIENCIAS, Colombia; MSMT CR, MPO CR and VSC CR, Czech
Republic; DNRF, DNSRC and Lundbeck Foundation, Denmark; ARTEMIS, European
Union; IN2P3-CNRS,\linebreak CEA-DSM/IRFU, France; GNAS, Georgia; BMBF, DFG, HGF, MPG
and AvH Foundation, Germany; GSRT, Greece; ISF, MINERVA, GIF, DIP and
Benoziyo Center, Israel; INFN, Italy; MEXT and JSPS, Japan; CNRST, Morocco;
FOM and NWO, Netherlands; RCN, Norway; MNiSW, Poland; GRICES and FCT,
Portugal; MERYS (MECTS), Romania; MES of Russia and ROSATOM, Russian
Federation; JINR; MSTD, Serbia; MSSR, Slovakia; ARRS and MVZT, Slovenia;
DST/NRF, South Africa; MICINN, Spain; SRC and Wallenberg Foundation,
Sweden; SER, SNSF and Cantons of Bern and Geneva, Switzerland; NSC, Taiwan;
TAEK, Turkey; STFC, the Royal Society and Leverhulme Trust, United Kingdom;
DOE and NSF, United States of America.

The crucial computing support from all WLCG partners is acknowledged
gratefully, in particular from CERN and the ATLAS Tier-1 facilities at
TRIUMF (Canada), NDGF (Denmark, Norway, Sweden), CC-IN2P3 (France),
KIT/GridKA (Germany), INFN-CNAF (Italy), NL-T1 \\ (Netherlands), PIC (Spain),
ASGC (Taiwan), RAL (UK) and BNL (USA) and in the Tier-2 facilities
worldwide.

\clearpage 
\onecolumn
\begin{flushleft}
{\Large The ATLAS Collaboration}

\bigskip

G.~Aad$^{\rm 48}$,
B.~Abbott$^{\rm 111}$,
J.~Abdallah$^{\rm 11}$,
A.A.~Abdelalim$^{\rm 49}$,
A.~Abdesselam$^{\rm 118}$,
O.~Abdinov$^{\rm 10}$,
B.~Abi$^{\rm 112}$,
M.~Abolins$^{\rm 88}$,
H.~Abramowicz$^{\rm 153}$,
H.~Abreu$^{\rm 115}$
% ...
\bigskip

% ...
$^{10}$ Institute of Physics, Azerbaijan Academy of Sciences, Baku, Azerbaijan\\
$^{11}$ Institut de F\'isica d'Altes Energies and Departament de F\'isica de la Universitat Aut\`onoma  de Barcelona and ICREA, Barcelona, Spain\\
% ...
$^{48}$ Fakult\"{a}t f\"{u}r Mathematik und Physik, Albert-Ludwigs-Universit\"{a}t, Freiburg i.Br., Germany\\
$^{49}$ Section de Physique, Universit\'e de Gen\`eve, Geneva, Switzerland\\
% ...
$^{88}$ Department of Physics and Astronomy, Michigan State University, East Lansing MI, United States of America\\
% ...
$^{111}$ Homer L. Dodge Department of Physics and Astronomy, University of Oklahoma, Norman OK, United States of America\\
$^{112}$ Department of Physics, Oklahoma State University, Stillwater OK, United States of America\\
% ...
$^{115}$ LAL, Univ. Paris-Sud and CNRS/IN2P3, Orsay, France\\
% ...
$^{118}$ Department of Physics, Oxford University, Oxford, United Kingdom\\
% ...
$^{153}$ Raymond and Beverly Sackler School of Physics and Astronomy, Tel Aviv University, Tel Aviv, Israel\\
% ...
\end{flushleft}



\begin{thebibliography}{10}

\bibitem{Chatrchyan2011123}
CMS Collaboration,
Phys.\ Lett.\ {\bf B704 } (2011)  123-142,
[arXiv:1107.4771 [hep-ex]].



\bibitem{Aad:2011aj}
  ATLAS Collaboration,
  New J.\ Phys.\  {\bf 13 } (2011)  053044,
  [arXiv:1103.3864 [hep-ex]].

\bibitem{Collaboration:2011fq}
  ATLAS Collaboration, 
  Phys.\  Lett.\ B (2011, submitted),
  [arXiv:1108.6311 [hep-ex]].

\bibitem{Aaltonen:2011sg}
  CDF Collaboration,
  Phys.\ Rev.\ Lett.\  {\bf 107 } (2011)  042001,
  [arXiv:1105.2815 [hep-ex]].

\bibitem{Chatrchyan:2011cj}
  CMS Collaboration,
  Phys.\ Rev.\ Lett.\ {\bf 107} (2011) 101801,
  [arXiv:1107.3084 [hep-ex]].

\bibitem{Kilic:2008ub}
  C.~Kilic, S.~Schumann, M.~Son,
  JHEP {\bf 0904 } (2009)  128,
  [arXiv:0810.5542 [hep-ph]].

\bibitem{PhenoOctet}
  S. Schumann {\it et al.},
  JHEP {\bf 09} (2011) 074,
 [arXiv:1108.2957 [hep-ph]].

\bibitem{Bai:2011mr}
  Y.~Bai, J.~Shelton,
  [arXiv:1107.3563 [hep-ph]].

\bibitem{Choi:2008ub}
  S.~Y.~Choi, M.~Drees, J.~Kalinowski, J.~M.~Kim, E.~Popenda, P.~M.~Zerwas,
  Phys.\ Lett.\  {\bf B672 } (2009)  246-252,
  [arXiv:0812.3586 [hep-ph]].

\bibitem{Kribs:2007ac}
  G.~D.~Kribs, E.~Poppitz, N.~Weiner,
  Phys.\ Rev.\  {\bf D78 } (2008)  055010,
  [arXiv:0712.2039 [hep-ph]].

\bibitem{Plehn:2008ae}
  T.~Plehn, T.~M.~P.~Tait,
  J.\ Phys.\ {\bf G36 } (2009)  075001,
  [arXiv:0810.3919 [hep-ph]].
  
\bibitem{Farrar:1978xj}
  G.~R.~Farrar, P.~Fayet,
  Phys.\ Lett.\  {\bf B76 } (1978)  575-579.

\bibitem{atlas_detector} ATLAS Collaboration, JINST {\bf 3}, (2008) S08003.


\bibitem{ATLAS-CONF-2010-094}
ATLAS Collaboration,
 ATLAS-CONF-2010-094, http://cdsweb.cern.ch/record/1299109.




\bibitem{Collaboration:2011tq}
ATLAS Collaboration, 
Eur.\ Phys.\ J.\ C (2011, accepted),
[arXiv:1107.2092 [hep-ex]].



\bibitem{Aad:1347441}
 ATLAS Collaboration,
 Phys.\ Rev.\ Lett.\ {\bf 107} (2011) 041802,
 [arXiv:1104.5225 [hep-ex]].


\bibitem{Aad:2011yb}
ATLAS Collaboration,
  Phys.\ Lett.\ B (2011, submitted),
  [arXiv:1108.3699 [hep-ex]].

\bibitem{Aad:1318461}
  ATLAS Collaboration,
  Phys.\ Lett.\  {\bf B698 } (2011)  325-345,
  [arXiv:1012.5382 [hep-ex]].




\bibitem{Mangano:2002ea}
  M.~L.~Mangano, M.~Moretti, F.~Piccinini, R.~Pittau and A.~D.~Polosa,
  JHEP {\bf 0307} (2003) 001,
  [arXiv:hep-ph/0206293].

\bibitem{Alwall:2007fs}
  J.~Alwall, S.~Hoche, F.~Krauss, N.~Lavesson, L.~Lonnblad, F.~Maltoni, M.~L.~Mangano, M.~Moretti {\it et al.},
  Eur.\ Phys.\ J.\  {\bf C53 } (2008)  473-500,
  [arXiv:0706.2569 [hep-ph]].

  
\bibitem{Corcella:2002jc}
  G.~Corcella {\it et al.},
  arXiv:hep-ph/0210213.

\bibitem{Butterworth:1996zw}
  J.~M.~Butterworth, J.~R.~Forshaw and M.~H.~Seymour,
  Z.\ Phys.\  C {\bf 72}, 637 (1996),
  [arXiv:hep-ph/9601371].

\bibitem{pythia}
  T.~Sjostrand, S.~Mrenna, P.~Z.~Skands,
  JHEP {\bf 0605 } (2006)  026,
  [hep-ph/0603175].

\bibitem{Sherstnev:2007nd}
  A.~Sherstnev and R.~S.~Thorne,
  Eur.\ Phys.\ J.\  C {\bf 55} (2008) 553,
  [arXiv:0711.2473 [hep-ph]].


\bibitem{Pumplin:2002vw}
  J.~Pumplin, D.~R.~Stump, J.~Huston, H.~L.~Lai, P.~M.~Nadolsky and W.~K.~Tung,
  JHEP {\bf 0207}, 012 (2002),
  [arXiv:hep-ph/0201195].

\bibitem{ATLAS-PHYS-PUB-2010-002}
ATLAS Collaboration,
ATLAS-PHYS-PUB-2010-002, http://cdsweb.cern.ch/record/1247375.

\bibitem{Agostinelli:2002hh}
  S.~Agostinelli {\it et al.} [ GEANT4 Collaboration ],
  Nucl.\ Instrum.\ Meth.\  {\bf A506 } (2003)  250-303.


\bibitem{cacciari-2008}
  M.~Cacciari, G.~P.~Salam, G.~Soyez,
  JHEP {\bf 0804 } (2008)  063,
  [arXiv:0802.1189 [hep-ph]].


\bibitem{ATLAS-CONF-2011-032}
ATLAS Collaboration,
ATLAS-CONF-2011-032, http://cdsweb.cern.ch/record/1337782.

\bibitem{Read:2002hq}
  A.~L.~Read,
  J.\ Phys.\ G {\bf G28 } (2002)  2693-2704.


\bibitem{Aad:2011dr}
  ATLAS Collaboration,
  Eur.\ Phys.\ J.\  {\bf C71 } (2011)  1630,
  [arXiv:1101.2185 [hep-ex]].







  
\end{thebibliography}
\end{document}